\documentclass[12pt,draft,notref,notcite]{article}

\usepackage{color,longtable}
\usepackage{amssymb,graphicx}

\newcommand{\be}{\begin{eqnarray}}
\newcommand{\ee}{\end{eqnarray}}

\usepackage{amsfonts,amsmath}
\usepackage{latexsym}

\def\cL{{\cal{L}}}

\def\cH{{\cal H}}

\def\cP{{\cal{P}}}
\def\cO{{\cal O}}
\def\cV{{\cal{V}}}

\def\a{{\alpha}}
\def\b{{\beta}}

\def\E{E_{10}}
\def\K{K(E_{10})}
\def\KE{E_{10}/K(E_{10})}

\def\ints{{\mathbb{Z}}}

\def\ni{\noindent}
\def\cL{{\cal L}}
\def\cV{{\cal V}}

\def\cP{{\cal P}}

\newcommand{\Ref}[1]{(\ref{#1})}

\begin{document}

\begin{center}
{\bf \Large Symmetries, Singularities and the\\[1mm] 
   De-emergence of Space}\\[7mm]
  Thibault Damour\footnotemark[1] and Hermann Nicolai\footnotemark[2]\\[3mm]
\footnotemark[1] {\sl Institut des Hautes Etudes Scientifiques\\
     35, Route de Chartres, F-91440 Bures-sur-Yvette, France}\\[1mm]
\footnotemark[2]{\sl  Max-Planck-Institut f\"ur Gravitationsphysik\\
     Albert-Einstein-Institut \\
     M\"uhlenberg 1, D-14476 Potsdam, Germany\\[1mm]
 Email: {\tt damour@ihes.fr, nicolai@aei.mpg.de}} \\[7mm]
\begin{minipage}{12cm}\footnotesize
\textbf{Abstract:} Recent work has revealed intriguing connections 
between a Belinsky-Khalatnikov-Lifshitz-type analysis of 
spacelike singularities in General
Relativity and certain infinite dimensional Lie algebras,
 and in particular the `maximally extended' hyperbolic Kac--Moody algebra $\E$.
In this essay we argue that these results may lead to an entirely new 
understanding of the (quantum) nature of space(-time) at the Planck 
scale, and hence -- via an effective `de-emergence' of space near 
a singularity -- to a novel mechanism for achieving background 
independence in quantum gravity.\\
\end{minipage}
\end{center}

{\bf 1. Introduction.}
A key challenge for a future theory of quantum gravity is the need to 
explain the fate of space-time singularities, where classical general 
relativity breaks down, and space and time `come to an end'. This challenge 
concerns in particular spacelike (cosmological) singularities, the most 
prominent example of which is the big bang singularity that gave birth 
to our universe. At issue here is not only the question of whether
and how quantum effects might {\em resolve} the singularity, but the very 
meaning of the term `singularity resolution' itself. The latter hinges 
essentially on what the correct theory is, and will almost certainly 
require new concepts that go beyond established notions of space and 
time.

A naive extension of quantum mechanics would suggest that 
singularity resolution works essentially in the same way for quantum 
general relativity as it does for the hydrogen atom. There, as is
well known, the expected classical `collapse' of an electron towards
the $1/r$ singularity of the Coulomb potential is resolved 
by the Heisenberg uncertainty principle and the quantum mechanical
smearing of the electron wave function, which allows the electron to
stay in a stable bound state around $r=0$. This mechanism is
often invoked in canonical approaches to quantum gravity, where
one would thus hope to be able to replace the classical time evolution 
of the spatial geometry, described as a `trajectory' in the space 
of 3-geometries (that is, Wheeler-DeWitt superspace), by a quantum 
mechanical description in terms of a wave functional which `smears' 
the 3-geometries over the singular classical trajectories. This line  
of thought has been extensively pursued in the simplified context of 
the mini-superspace approximation, with varying results: while models 
derived from (or motivated by) loop quantum gravity generally tend 
to predict a `bounce' providing a quantum mechanical bridge between 
two classical universes~\cite{Bojo}, the more conventional quantum 
geometrodynamical treatment of the mini-superspace Wheeler-DeWitt 
equation shows no such evidence \cite{Kiefer}.

In this essay we would like to outline a very different proposal,
motivated by recent work \cite{DaHeNi02,DaNi05}, where the singularity 
is `resolved' by the effective `disappearance' of space, and the 
replacement of the dynamical fields, most notably the spatial metric 
$ g_{ij}(t,{\bf x})$, by a single dynamical variable $\cV(t)$ belonging
to an {\em infinite dimensional} coset space and depending only on time.
Our proposal is based on the discovery of a profound relation between 
an analysis {\it \`a la} Belinsky-Khalatnikov-Lifshitz (BKL) of 
spacelike singularities \cite{BKL,Misner} on the one hand, and the 
theory of indefinite Kac-Moody algebras on the other \cite{DaHe01,DaHeNi03} 
(see \cite{Kac} for an introduction to the theory of Kac--Moody algebras). 
More specifically, the main conjecture of \cite{DaHeNi02}, formulated 
in the context of the maximally extended $D=11$ supergravity \cite{CJS}, 
relates a BKL-type expansion in spatial gradients at a given spatial 
point to a Lie algebraic expansion in the height of certain roots of the 
`maximally extended' hyperbolic Kac--Moody algebra $\E$. Thereby the 
time evolution of $10$-dimensional geometric data is mapped onto an 
{\em effectively one-dimensional dynamics}, namely a (constrained) null 
geodesic motion in the infinite dimensional coset space $\KE$, which is 
formally defined as the quotient of the group $\E$ by its maximal 
compact subgroup $K(\E)$~\footnote{The emergence of $\E$ in the dimensional 
 reduction of maximal supergravity to one dimension had been conjectured 
 already long ago \cite{Ju85}. See also \cite{West} for a conceptually 
 very different proposal based on $E_{11}$.}. 
The appearance of $\E$ in this context is both unexpected and remarkable,
because $\E$ enjoys a similarly distinguished status among the
infinite dimensional Lie groups~\footnote{For simplicity of notation, 
we denote the group and its Lie algebra by the same symbol.} as the 
exceptional group $E_8$ does in the Cartan-Killing classification 
of the  finite dimensional simple Lie groups \cite{E8}.

\vspace*{0.4cm}
\ni
{\bf 2. BKL and cosmobilliards.}
We start by summarizing the BKL-type analysis of the {\em near
spacelike singularity limit}, that is, of the asymptotic behaviour 
of various fields, and in particular the (spatial) metric, near a 
singular hypersurface, here taken to be `located' at $T=0$ in proper
time $T$. To this aim, it is convenient \cite {DaHeNi03} to decompose 
the $D=11$ metric $g_{\mu \nu}$ into non-dynamical and dynamical
components, namely the lapse $N$ and the shift vector $N^m$ (set here 
to zero), and into `diagonal' and `off-diagonal' components $e^{- 2\b^a}$ 
and $\theta^a{}_i$, respectively, such that the line element becomes
\be
ds^2 = - N^2 dt^2 + 
\sum_{a=1}^{10} e^{- 2\b^a} \theta^a{}_i \theta^a{}_j  dx^i dx^j
\ee
Here, the `off diagonal' components $\theta^a{}_i$ entering
the Iwasawa decomposition of the spatial metric $g_{ij}$ 
are {\em upper triangular matrices} with 1's on the diagonal.
We choose a gauge for $N$ in terms of $g_{ij}$ in such a way that 
$N \sim \cO(T) \to 0$ when $T\rightarrow 0$. Thus $t\sim - \log T$
becomes a `Zeno-like' time coordinate with $t\rightarrow +\infty$ as 
$T\rightarrow 0$.

The Hamiltonian constraint, at a given spatial point, can be written as
\be \label{V4}
\cH(\b^a, \pi_{a},Q,P) = \tilde{N} \left[\frac12 G^{ab} \pi_a \pi_b  +
\sum_A c_A (Q,P,\partial\b,\partial Q)\exp\big(- 2 w_A (\b)\big)\right]
\ee
with the rescaled lapse $\tilde{N} \equiv N/ \sqrt{g}$, where $g$ is 
the determinant of the spatial metric. Here $\pi_a$ (with $a=1, \dots, 10$) 
are the canonical momenta conjugate to the logarithmic scale factors $\b^a$, 
and $G^{ab}$ is the (Lorentzian) DeWitt `superspace' metric induced 
by the Einstein-Hilbert action. $(Q,P)$ denote the remaining canonical
degrees of freedom associated to the off-diagonal metric components 
$\theta^a{}_i$ and various matter degrees of freedom [such as the 
3-form $A_{\mu \nu \lambda}(t,{\bf x})$ of $D=11$ supergravity], 
as well as their respective conjugate momenta, and 
$(\partial\b, \partial Q)$ are the {\em spatial}  gradients of $\b$ 
and $Q$. The exponential terms in \Ref{V4} involve {\em linear forms} 
$w_A(\b) \equiv w_{A a}  \b^a$, where the specific coefficients 
$ w_{A a}$ and the range of labels $A$ depend on the model under 
consideration (see \cite{DaHeNi03} for details). 

The BKL limit $T\to 0$ amounts to considering the {\em large $\b$ limit}
in Eq.~(\ref{V4}), and is determined by the exponential `walls' 
$\propto \exp (- 2 w_A (\b))$ \cite{DaHeNi03}. The latter can be ordered 
in `layers'. The first layer, corresponding to the subset of `dominant 
walls' $w_{A'}(\beta)$ --- whose coefficients 
$c_{A'}(Q,P,\partial\b,\partial Q)$ can be shown to be 
non-negative --- confine the motion in $\b$-space to a {\em fundamental
billiard chamber} defined by the inequalities $w_{A'}(\beta) \geq 0$. 
The remaining (sub-dominant) exponential walls introduce fractional 
corrections to the chaotic motion of $(\b^a, \pi_{a})$ within the 
fundamental billiard chamber. All the other dynamical variables  
$(Q,P)$, together with their spatial gradients, `freeze' as $T\rightarrow 0$, 
and thus exhibit a very different behavior in this limit.

\vspace*{0.4cm}
\ni{\bf 3. Coset space dynamics.} 
Let us next consider an {\it a priori} very different dynamical
system, namely null geodesic motion on the infinite-dimensional 
coset space $\KE$. A curve on this coset space can be parametrized
by a time-dependent (but {\em space independent}) element of the
$\E$ group in upper triangular (Iwasawa) form: $\cV(t) = \exp h(t) 
\exp \nu(t)$. Here, $h(t) = \b^a(t) H_a$ belongs to the 10-dimensional 
Cartan subalgebra (= CSA) of $\E$. Our use of the same notation as
above is justified by the fact that we will eventually identify the ten 
CSA coordinates $\b^a$ of $\E$ with the logarithms of the ten `diagonal' 
components of the spatial metric $g_{ij}$ introduced above. On the 
other hand, $\nu(t) = \sum_{\a >0} \nu^{\a}(t) E_{\a}$
belongs to a (Borel) subalgebra of $\E$ and has an infinite number
of components labelled by positive roots $\a$ of $\E$. The geodesic
action is formally very simple; it reads 
\be\label{L}
\int dt \, \cL(t) = \int \frac{dt}{n(t)} \, \langle \cP(t)|\cP(t)\rangle\; ,
\ee
where $\cP := (1/2)\big[\dot\cV \cV^{-1} + (\dot\cV \cV^{-1})^T\big]$ is 
the `velocity'\footnote{Here the transpose $^T$ denotes the negative 
 of the Chevalley (`compact') involution \cite{Kac}. The compact subalgebra 
 $\K$ is thus spanned by the `antisymmetric' elements of $\E$.}, 
$\langle .|.\rangle$  is the standard invariant bilinear form 
generalizing the finite dimensional matrix trace \cite{Kac}, and $n(t)$ 
is a one-dimensional `lapse' needed to ensure (time) reparametrisation 
invariance of the action \Ref{L}. The Zeno-like coordinate time $t$ of 
the previous section is recovered upon identifying $n$ with the rescaled 
lapse $\tilde{N}$ introduced after \Ref{V4} and choosing the gauge $n(t)=1$.

Varying (\ref{L}) w.r.t. to the lapse $n$, we obtain the Hamiltonian 
constraint:
\be\label{HKM}
H(\b^a,\pi_a, \nu, p) =  n  \left[  \frac12  G^{ab} \pi_a \pi_b
      + \sum_{\a > 0} \sum_{s=1}^{{\rm mult}(\a)}
      \big(\Pi_{\a ,s}(\nu^{\a},p_{\a})\big)^2 \exp\big(-2\a(\b)\big) \right]
\ee
where $\pi_a$  denote the conjugate momenta of the ten diagonal 
CSA coordinates $\b^a$, and $ p_{\a}$ denote the conjugate momenta 
of the `off-diagonal' coordinates $\nu^{\a}$ parametrizing the Borel 
part of $\cV$, on which the $\Pi_{\a,s}$ depend \cite{DaHeNi03}. 
The sum on the r.h.s. of \Ref{HKM} ranges over all positive 
roots $\a$ of $E_{10}$ with their multiplicities [$={\rm mult}(\a)$]. 
We recall that the roots $\a$ are {\em linear forms} on the CSA, that
is, we have  $\a(\b) \equiv \a_a \b^a$ for the exponents in \Ref{HKM}.
Although the dynamics encapsulated in \Ref{HKM} is very complicated,
a general feature is that in order to satisfy $H=0$, we must always have 
$G^{ab} \pi_a \pi_b \leq 0$; this means that the coset null geodesics 
must always maintain a future-directed CSA velocity $\pi^a$, hence 
{\em cannot bounce} backwards in $\b$ space.

\vspace*{0.4cm}
\ni{\bf 4. Correspondence between BKL and coset-space dynamics.}
The formal similarity between the gravitational Hamiltonian \Ref{V4} 
(considered at a given spatial point) and the coset Hamiltonian \Ref{HKM} 
is evident, but the precise correspondence has so far been established
only for a limited number of terms. In particular, the metric $G^{ab}$ 
entering \Ref{HKM}, which is the restriction of the invariant bilinear 
form on $E_{10}$ to its CSA, happens to be {\em identical} with the DeWitt 
metric appearing in \Ref{V4}. This fact enables us to {\em identify 
the space of logarithmic scale factors with the Cartan subalgebra of $\E$} 
(as anticipated by our notation).
Moreover, one can analyze the asymptotic dynamics of the 
coset variables $(\b^a, \pi_a, \nu^{\a}, p_{\a})$ in 
the limit of large $\b$'s. At first order in an expansion in `height' 
of the simple roots of $\E$, one finds that the CSA variables $\b$
are confined to a chaotic billiard motion within the Weyl chamber 
of $\E$. The latter is defined by the inequalities $\a_i(\b) \geq 0$, 
where the $\a_i$'s ($i=1, \cdots, 10$) are the {\em simple roots} of 
$\E$, and turns out to {\em coincide} with the fundamental BKL billiard 
chamber defined by the dominant potential walls $w_{A'}(\beta)$ for
$D=11$ supergravity. Consideration of the subleading exponential walls 
in both models now shows that one can actually identify the 
two dynamics {\em up to height 30}, i.e. much beyond the leading 
billiard dynamics (corresponding to height one only) \cite{DaHeNi02}. 
This result suggests the existence of a {\em hidden equivalence} between 
the two models, i.e. the existence of a map preserving the dynamics
between the infinite tower of coset variables $(\b^a, \pi_a, \nu^{\a}, 
p_{\a})$, and the infinite sequence of spatial Taylor coefficients  
$(\b({\bf x}),\pi({\bf x}), Q({\bf x}),P({\bf x}), 
\partial \b({\bf x}), \partial Q({\bf x}), \partial^2 \b({\bf x}), 
\partial^2 Q({\bf x}), \dots)$  formally describing the dynamics of 
the (super)gravity fields $(\b({\bf x}),\pi({\bf x}), Q({\bf x}),
P({\bf x}))$ in the neighborhood of some given spatial point ${\bf x}$. 
In this way, a skeletonization of the (super-)gravity fields by means 
of their infinite towers of spatial gradients  gets related to a purely 
Lie algebraic expansion in terms of heights of roots. While the full 
details of this correspondence (which is expected to be ultimately very 
non-local in the space-time fields) remain to be worked out, it has 
been possible recently to extend these results also to the fermionic 
sector on both sides \cite{DKN06,dBHP06}.
  
Most importantly for our present proposal, certain (partially) known 
{\em quantum corrections} to the classical supergravity action can be
shown to be compatible with specific terms, of very large height, present 
in the coset action \cite{DaNi05}. For instance, the leading term 
quartic in the Weyl curvature, 
 \be\label{C4}
\cL^{(4)} &=& 192 N\sqrt{g}
\Big(- C^{ABCD} {C_{AB}}^{EF} {C_{CE}}^{GH} C_{DFGH} \nonumber\\
&& \qquad\quad + \,4 \, C^{ABCD} C_A{}^E{}_C{}^F 
    C_E{}^G{}_B{}^H C_{FGDH}\Big)
\ee
is dominated, near the singularity, by an exponential term 
$\propto \exp[ - 2 \a(\b)]$ in the coset Hamiltonian \Ref{HKM} 
for a specific {\em imaginary} $\E$ root $\a$ of height (minus) $115$. 
Detailed study of the combination of curvature terms in \Ref{C4} 
has established an inequality \cite{DaNi05} confirming the {\em no bounce} 
property (in $\b$ space) exhibited by the coset dynamics, and briefly
explained after \Ref{HKM}.

\vspace*{0.4cm}
\ni{\bf 5. The cosmological singularity: a new paradigm?} 
The evidence summarized above suggests an entirely new picture of the 
(quantum) fate of space and time at a cosmological singularity. Namely 
we here propose to take seriously the idea that near the singularity (i.e. 
when the curvature gets larger than the Planck scale) the description 
in terms of a spatial continuum and space-time based (quantum) field
theory breaks down, and should be replaced by a much more abstract 
Lie algebraic description. Thereby the information previously encoded 
in the spatial variation of the geometry and of the matter fields gets 
transferred to an infinite tower of Lie algebraic variables depending 
only on `time'. In other words,  we are led to the conclusion that 
space --- and thus, upon quantization, also space-time --- actually
{\em disappears} (or `de-emerges') as the singularity is approached.
\footnote{We have in mind here a `big crunch', i.e.  we conventionally
consider that we move  
  {\em towards} the singularity. {\it Mutatis 
 mutandis}, we would say that space `appears' or `emerges' at a big bang.} 
There is no `quantum bounce' bridging the gap between an incoming 
collapsing and an outgoing expanding quasi-classical universe. 
Instead `life continues' at the singularity for an {\em infinite 
affine time}, however, with the understanding that $(i)$ dynamics no longer 
`takes place' in space, and $(ii)$ the infinite affine time interval 
[measured, say, by the Zeno-like time coordinate $t$ of \Ref{L}] corresponds 
to a sub-Planckian interval $0< T< T_{\rm Planck} $ of geometrical 
proper time.

Upon quantization, the geodesic equations of motion following from \Ref{L}
are replaced by a quantum version of the Hamiltonian constraint \Ref{HKM}, 
analogous to the Wheeler-DeWitt equation, and acting on some
`wave function of the coset particle' $\Psi = \Psi (\b^a,\nu^\alpha)$
depending on the coset variables. This, then, is the step where time 
also `disappears': as in all canonical approaches to quantum gravity, 
the wave function (or functional) $\Psi$ no longer depends on any 
extrinsic `time' [although one can, of course, choose a `clock field' 
among the coset variables so as to define an `operational' time, in terms 
of which the quantum dynamics of the remaining variables can be 
parametrized]. The quantum constraint would take the form of a 
Klein-Gordon-like equation~\footnote{Or a  `Dirac-like' (first order) 
constraint if fermions are included \cite{DKN06,dBHP06}.} 
\be \label{KG}
\Box \Psi(\b^a, \nu^{\a}) =0
\ee
where $\Box$ is the (formal) Laplace-Beltrami operator on the 
infinite-dimensional (Lorentzian) coset manifold $\KE$. It is
noteworthy that {\em all reference to space and time
has disappeared} in \Ref{KG}. The discretization of 
finite-dimensional duality symmetries upon quantization, well known
from toroidal compactification in string theory, would then suggest 
that the `wave function of the universe' is a modular form over 
the arithmetic group $\E(\ints)$ \cite{Ganor}.

\vspace*{0.4cm}
\ni{\bf 6. Outlook.} 
If correct, the picture outlined here will not only affect our 
understanding of what `happens' at a cosmological singularity, 
but may also shed a completely new light on the issue of background 
independence in quantum gravity. More succinctly, taking the quantum 
coset dynamics \Ref{KG} as a guiding principle, the correct theory 
of quantum gravity may well turn out to be background independent in 
the sense that near the singularity, the theory --- rather than
`quantizing' the spatial geometry, or some other spatially extended 
background structure --- simply does away with the background altogether, 
whence the whole issue would become moot!

Let us also note some potentially important implications of this
picture for the so-called `information loss paradox' in black hole 
physics. Indeed, the present ideas might also be applied to the
case of a `localized' big crunch (as the one inside a  black hole formed
within an asymptotically flat space-time). It would then suggest that
some of the information contained within the horizon might transmigrate
to the state of motion of a coset `baby particle' (whose dynamics
describes physics at sub-Planckian scales near the big crunch). The 
Hawking evaporation of the black hole containing this localized big 
crunch poses interesting conceptual challenges with regard to an 
infinite affine `coset life-time' near the singularity.

It is well known that symmetry concepts have been of central 
importance in the advancement of theoretical physics over the last 
century. They have been a key ingredient in the development of the 
two most successful theories of physics, namely general relativity 
(via the principle of general covariance) and the standard model 
of elementary particle physics (via gauge invariance and Yang Mills 
quantum field theories). In view of its distinguished place among 
all Lie algebras, $\E$ is a most worthy candidate for symmetry of 
nature, {\em deeply intertwining space-time with matter degrees of freedom}, 
and thus necessarily implying a unification of gravity and matter.
For this reason, we can anticipate for it a key role in elucidating 
the quantum nature of space-time, and hence space-time singularities.

\vspace{0.2cm}\ni
{\bf Acknowledgements}: It is a pleasure to thank our collaborators
Marc Henneaux and Axel Kleinschmidt for many fruitful discussions 
over the years. H.N. would also like to thank KITP, Santa Barbara, 
for hospitality in January 2007, and the workshop participants for 
discussions and inspiration.

\end{document}